# Crescent domain-wall pairs in anisotropic two-dimensional MnOI monolayer with fast dynamics

Yijun Yang, Zhong Shen, Jun Chen, Shuai Dong, and Xiaoyan Yao[*]

*Key Laboratory of Quantum Materials and Devices of Ministry of Education,*

*School of Physics, Southeast University, Nanjing 211189, China*

[*]Email: yaoxiaoyan@seu.edu.cn

**Abstract:** The controllable and efficient manipulation is always a key challenge for the application of topological magnetic textures in spintronic devices. By first-principles calculations and atomistic simulations, the present work reveals an exotic domain-wall pair in crescent shape, which possesses topology in one dimension and particle-like robustness in two dimensions. Its antiferromagnetic version is observed in the two-dimensional MnOI monolayer with a strong anisotropy, and the ferromagnetic version exists under an appropriate strain. Both of them can be driven efficiently by current to move in a straight line along a certain direction. Hereby, another possible path is provided to realize the application of topological magnetic texture in the next-generation high-speed spintronic devices.

# Introduction

In recent years, the topologically nontrivial magnetic textures have attracted a great deal of interest in both theoretical and experimental fields, due to their topologically protected stability and potential applications for nonvolatile energy-efficient spintronic devices, such as the random access memory and logic devices [1,2]. Different from the singular topological point defects which require a vanishing magnetization at the center, the smooth topological textures have a smooth magnetization everywhere, resembling 'knots' in the magnetization field that cannot continuously be transformed to the uniform magnetic state (e.g. ferromagnetic state) [3]. Among these, the coplanar domain wall is the simplest one-dimensional (1D) paradigm [1,4], and skyrmion is a much highlighted representative of two-dimensional (2D) cases [5-9]. Hopfion belongs to the case of three dimensions, which is more complex and has been studied in recent years [10-12].

For the potential application of topological magnetic textures as information carriers in spintronic devices, the essential requirement is the controllable motion, which includes two key aspects: the maximum velocity and the fixed direction. There are many driving methods such as magnetic field, spin-polarized current, voltage and so on [13-16]. Meanwhile, there are also many obstacles to achieve controllable fast motion. For domain walls, the nanowires or artificial narrow bands are usually required to realize manipulation, but usually there is strong pinning at edges. For skyrmions, their dynamics are often perturbed by the skyrmion Hall effect [17]. Up to now, the velocity in most cases was limited to a few hundred meters per second [18-22]. To efficiently manipulate stable topological magnetic textures with high velocity is still one technological and scientific key challenge.

Rising to the challenge, one approach is to improve domain wall or skyrmion. Antiferromagnets without net magnetization, which are difficult to detect but produce zero stray fields and ultrafast magnetization dynamics [23], could be such a candidate. Experimentally it was reported that the domain wall was moved efficiently by current pulses in synthetic antiferromagnetic racetracks at a speed of 750 m/s [24], and skyrmions in compensated synthetic antiferromagnets were moved by current without skyrmion Hall effect at speeds of up to 900 m/s [25]. Moreover, in compensated ferrimagnets the speed of a current-driven domain wall motion reached 1.3 km/s [26]. On the other hand, some theoretical works predicted that the domain walls and skyrmions in the antiferromgnetic (AFM) cases can move at speeds exceeding 1 km/s [27,28].

To meet the challenge, another way is to explore new topological magnetic texture and the corresponding dynamics. In this work, a 2D monolayer MnOI is predicted by the first-principles calculations. The strong anisotropy with a sizeable Dzyaloshinskii-Moriya interaction (DMI) is observed due to its anisotropic Janus structure. A transition from the AFM to the ferromagnetic (FM) ground state can be realized by applying strain. The atomistic simulation reveals that exotic AFM and FM domain-wall pairs in crescent shape exist as metastable states. Different from 1D domain wall and 2D skyrmion, these crescent domain-wall pairs (CDWPs) possess topology in one dimension, and particle-like nature in two dimensions. These CDWPs show interesting dynamic behaviors driven by magnetic field and spin-polarized current. Due to the strong anisotropy of structure, their motions are strictly limited along the direction of easy-axis. Moreover, with high stability, the straight motion could reach a high speed on the 2D plane.

## Methods

The first-principles calculations are performed with the Vienna *ab initio* Simulation Package (VASP), based on the density functional theory (DFT) [29,30]. The exchange-correlation potential is characterized by the Perdew-Burke-Ernzerhof (PBE) of the generalized gradient approximation (GGA) [31]. To better describe the on-site Coulomb interaction of localized 3$d$ states of Mn, an effective Hubbard term $U$ is set to be 3 eV [32,33]. The projector-augmented wave (PAW) method is adopted, and the plane-wave cutoff energy is set as 420 eV. To sample the Brillouin zone, we use $19 \times 19 \times 1$ Γ-centered k-point meshes for the primitive cell. To obtain accurate parameters, we set a high convergence standard with the energy and force less than $10^{-7}$ eV and 0.001 eV/Å, respectively. A vacuum space of 17 Å is adopted to avoid the interactions between the adjacent layers along the $z$ direction. Based on density functional perturbation theory (DFPT) [34], the phonon spectrum is calculated on a $5 \times 5 \times 1$ supercell by using the PHONOPY code [35].

To explore the spin textures and the spin dynamics based on the magnetic parameters obtained from first-principles calculations, we perform atomistic simulations with the Heisenberg model and Landau-Lifshitz-Gilbert (LLG) equation [36] as implemented in the SPIRIT package [37]. To achieve a stable state, a $100 \times 100 \times 1$ supercell with periodic boundary conditions, $2 \times 10^5$ iterations and 0.001 ps for time step are chosen. When the dynamics is explored, the initial state is the state after full relaxation of a $400 \times 10 \times 1$ supercell with a single CDWP on the ground state.

## Results and discussion

### A. Structure and magnetism

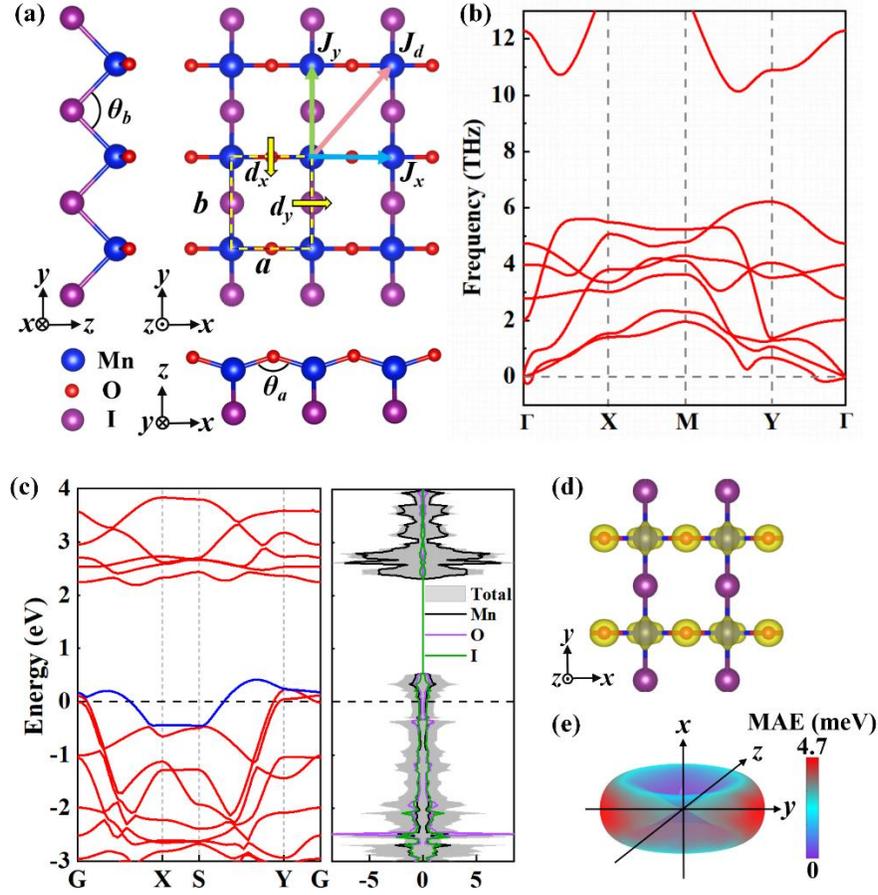

**FIG. 1.** (a) Crystal structure of MnOI monolayer. Blue, red, and purple balls represent Mn, O and I, respectively. The blue, green, pink and yellow arrows in the top view indicate the exchange couplings ($J_x$, $J_y$, $J_d$) and in-plane components of DMI ($d_x$, $d_y$). (b) Phonon spectrum. (c) The band structure and density of states of MnOI monolayer. (d) The partial electronic density distribution of the band crossing the Fermi level (the blue band in Fig. 1(c)). (e) Angular dependence of calculated MAE on spin orientation, where the MAE is set to zero in the direction of $x$ axis.

The structure optimization demonstrates that the Janus MnOI monolayer has an orthorhombic structure with the space group *Pmm*2 (No. 25) and broken spatial inversion symmetry. As shown in Fig. 1(a), each Mn atom forms bonds with two O atoms in the $x$ direction and two I atoms in the $y$ direction. The bond angles of Mn-O-Mn ($\theta_a$) and Mn-I-Mn ($\theta_b$) are 144.582° and 92.462°, respectively. The dashed lines show the unit cell of MnOI with the optimized lattice constants $a$ = 3.491 Å and $b$ = 4.019 Å.

To verify the stability of MnOI monolayer, the phonon spectrum is calculated. As shown in Fig. 1(b), except the negligible imaginary frequency at about Γ-point which also appeared in other stable two-dimensional materials [38-41], no other imaginary frequency can be observed. Therefore, the structure is dynamically stable. We also calculate the average cohesive energy ($E_{coh}$) to evaluate the thermodynamical stability.

$$E_{coh} = (E_{Mn} + E_O + E_I - E_{total})/3 \tag{1}$$

where $E_{total}$ is the energy of monolayer MnOI. $E_{Mn}$, $E_o$, and $E_I$ are the energies of single Mn, O, and I atom, respectively. The obtained $E_{coh}$ is 3.30 eV/atom, which is comparable to other 2D monolayers like $Fe_2I_2$ (3.01 eV/atom) [42] and $Cu_2Si$ (3.46 eV/atom) [43]. Based on the experimental synthesis strategy of Janus monolayer [44-47] or criteria of 'easily exfoliable' [48], the MnOI monolayer could be experimentally accessible (see Note 3(a) in SM [49] for details).

As illustrated in Fig. 1(c), the band structure and density of states demonstrate the metallicity of MnOI monolayer. Around Fermi level, the $d$ orbitals of Mn are strongly hybridized with the $p$ orbitals of O and I. Fig. 1(d) depicts the partial electronic density distribution of the band crossing the Fermi level (the blue band in Fig. 1(c)). It is seen that the electrons mostly stay on the shorter bond along the $x$ direction, which induce much stronger interactions along the $x$ direction.

**Table I.** Magnetic moment of Mn ($m$), exchange coupling coefficients ($J_x$, $J_y$, $J_d$), in-plane DMI components ($d_x$, $d_y$) and magnetic anisotropy coefficient ($K$). The unit of $m$ is $\mu_B$. The units of $J_x$, $J_y$, $J_d$, $d_x$, $d_y$ and $K$ are meV.

| $m$ | $J_x$ | $J_y$ | $J_d$ | $d_x$ | $d_y$ | $K$ |
|---|---|---|---|---|---|---|
| 4.05 | 62.4 | -2.7 | -0.6 | -8.8 | -1.2 | -4.7 |

The magnetic moment of Mn ($\mu_{Mn}$) is 4.05 $\mu_B$. The calculation on the magnetic anisotropy energy (MAE) demonstrates that this monolayer shows a strong easy-axis anisotropy along the $x$ direction, and is approximately isotropic in the $yz$ plane, as illustrated in Fig. 1(e). To investigate magnetic properties, the following spin Hamiltonian can be applied.

$$\mathbf{H}_{spin} = -J_{ij}\sum_{\langle i,j \rangle} \mathbf{S}_i \cdot \mathbf{S}_j + K\sum_i (S_i^x)^2 - \sum_{\langle i,j \rangle} \mathbf{D}_{ij} \cdot (\mathbf{S}_i \times \mathbf{S}_j) - \mu_{Mn} h \sum_i S_i^x \tag{2}$$

where $\mathbf{S}_i$ is the normalized spin vector of $i_{th}$ Mn atom, and $<i, j>$ denotes the summation over all the neighboring Mn pairs. $J_{ij}$ represents the exchange coupling between the neighboring $S_i$ and $S_j$, that is, $J_x$, $J_y$ and $J_d$ in the $x$, $y$, and diagonal directions, as marked in Fig. 1(a). $K$ is the magnetic anisotropy coefficient, which is calculated based on MAE (see Note 1 in SM [49] for details). According to the structure symmetry, for each Mn-Mn bond, there is always a mirror plane including it, and then the DMI vector $\mathbf{D}_{ij}$ is perpendicular to this mirror plane according to Moriya's rules [50], namely $\mathbf{D}_{ij} = (0, d_x, 0)$ at the Mn-Mn bond along the $x$ direction and $\mathbf{D}_{ij} = (d_y, 0, 0)$ at that along the $y$ direction, as illustrated by the yellow arrows in Fig. 1(a). $h$ represents the external magnetic field, which is applied along the easy-axis. The dipolar interactions are neglected because they hardly affect the AFM structure (see Note 2(c) in SM [49] for details).

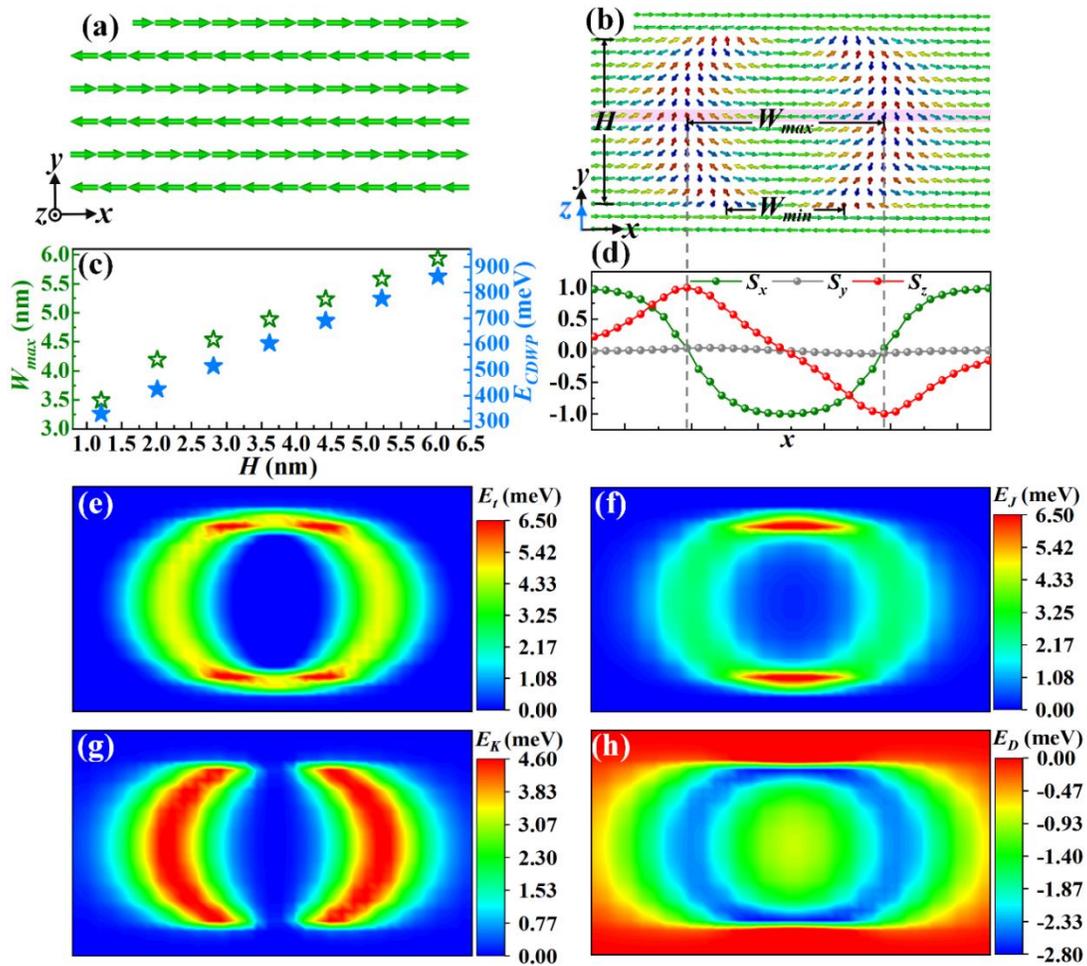

**FIG. 2.** Snapshots of (a) XStripy-AFM state and (b) AFM CDWP, where its width ($W_{max}$, $W_{min}$) and height ($H$) are estimated by the marked distances. (c) $W_{max}$ and the energy of one CDWP ($E_{CDWP}$) depend on $H$. (d) The $x$, $y$ and $z$ components ($S_x$, $S_y$, $S_z$) of the spins outlined by a pink shadow in (b). The distributions of (e) the total energy ($E_t$), (f) the energy of exchange interactions ($E_J$), (g) the energy of anisotropy ($E_K$), and (h) the energy of DMI

($E_D$) for an isolated AFM CDWP, where the ground state (XStripy-AFM) is taken as the reference.

By mapping energies of typical magnetic orders to the Hamiltonian (Eq. 2), the magnetic parameters of MnOI monolayer are obtained, which are listed in Table I. In addition, the perturbation method is applied to calculate the longer-range exchange interactions [51,52], which have been confirmed not to affect the main results. (see Note 1(a) in SM [49] for details). Due to the strong easy-axis anisotropy with a large $K$, all the magnetic moments of Mn tend to orient along the $x$ axis. The FM $J_x$ (62.4 meV) is far stronger than the AFM $J_y$ (-2.7 meV) and $J_d$ (-0.6 meV), leading to the XStripy-AFM state as ground state, where the $x$-direction FM strips are antiferromagnetically coupled in the $y$ direction, as illustrated in Fig. 2(a). The DFT calculation on typical magnetic orders also confirms that XStripy-AFM state owns the lowest energy. Based on the anisotropic Janus structure, DMI also exhibits a strong anisotropy, that is, $d_x$ (8.8meV) is much larger than $d_y$ (1.2meV), indicating the spin spiral with one dominant direction.

Based on the magnetic parameters obtained from the DFT calculations, the atomistic simulations are performed, which confirm that the ground state is XStripy-AFM state, as shown in Fig. 2(a). It is interesting that an exotic AFM crescent domain-wall pair (CDWP) exists as metastable state without external magnetic field, as displayed in Fig. 2(b). The CDWP possesses a localized particle-like profile, which can be characterized by the width ($W_{max}$, $W_{min}$) and the height ($H$) as marked on Fig. 2(b). It consists of two 180° AFM Néel walls with a certain chirality determined by DMI. These two parts are tightly bonded with a fixed $W_{min}$ of about 3.1 nm. $W_{max}$ depends on $H$ approximately linearly, as plotted in Fig. 2(c). Both height and width are only several nano meters, and thus the CDWP is a nanoscale magnetic texture on two dimensions.

Taking the ground state as the reference, the energy of a single AFM CDMP ($E_{CDWP}$) depends on $H$ approximately linearly, as plotted in Fig. 2(c). The values are comparable to those experimentally accessible metastable states, such as domain walls and skyrmions [53-56]. At the same time, the energy barrier between a single AFM CDWP and the ground state is higher than some energy barriers reported for the annihilation of skyrmions [54,55]. Therefore, these CDWPs could be experimentally accessed (see Note 3(c) in SM [49] for details).

To analyze the causes of the crescent shape, the energy distributions are presented in Fig. 2(e-h). It is seen that the top and bottom interfaces of CDWP show higher total energy ($E_t$) (Fig. 2(e)). To reduce $E_t$, these interfaces should be shortened, and therefore the domain walls bend inward and the crescent shape is formed. The detailed calculation indicates that this shape results from the competition between the exchange interactions, DMI and anisotropy, in which the exchange interactions make dominant contribution.

Along the easy-axis in the $x$ direction, the strong $d_x$ produces a spin spiral with 360° rotation within the $xz$ plane as shown in Fig. 2(d). Since the spin space is confined to two dimensions by the strong $K$ and $d_x$, a 1D topological magnetic structure is formed. Its topology can be characterized by the 1D soliton winding number calculated as [57,58]:

$$w = \frac{1}{2\pi} \int_{x_1}^{x_2} \partial_x \phi \, dx \tag{3}$$

where $x_1$ and $x_2$ are positions deep inside the domains on two sides of soliton. $\phi$ is the angle that the spin makes with the easy-axis. Here, $w = 1$ is obtained. Therefore, the nanoscale CDWP can be regarded as a 2D particle-like magnetic texture with 1D topology.

## B. Strain modulation

As a conventional and powerful modulation on flexible 2D materials, strain is applied on both $x$ and $y$ directions to investigate the variation of magnetic properties. Here, the strength of in-plane biaxial strain is defined as $\varepsilon = \dfrac{a - a_0}{a_0}\%$, where $a$ and $a_0$ represent the in-plane lattice constants of strained and pristine monolayers. $\varepsilon$ is ranged from compressive -3% (~0.4 GPa) to tensile 4% (~ -0.9 GPa), which is achievable in experiment. When the strain is applied by changing the in-plane lattice constants, the structure is optimized and the atomic positions are adjusted with the lattice constants fixed. Then the magnetic parameters are extracted from the new energies after optimization. Fig. 3(a) plots the energy difference between FM and XStripy-AFM states ($E_{FM}$-$E_{XStripy-AFM}$) as a function of $\varepsilon$. (The other AFM states with much higher energies are not shown here.) It is seen that a transition of ground state occurs at a strain of ~1.8%. Below ~1.8%, XStripy-AFM configuration is preferred. Above ~1.8%, FM state is favored. Correspondingly, the dependences of magnetic parameters on strain are plotted in Fig. 3(a-b). Although $J_x$ declines slowly with the increasing $\varepsilon$, it maintains the dominant position on the whole range of $\varepsilon$. $J_d$ undergoes a transition from negative to positive values, which plays a key role in the ground state transition from XStripy-AFM to FM state.

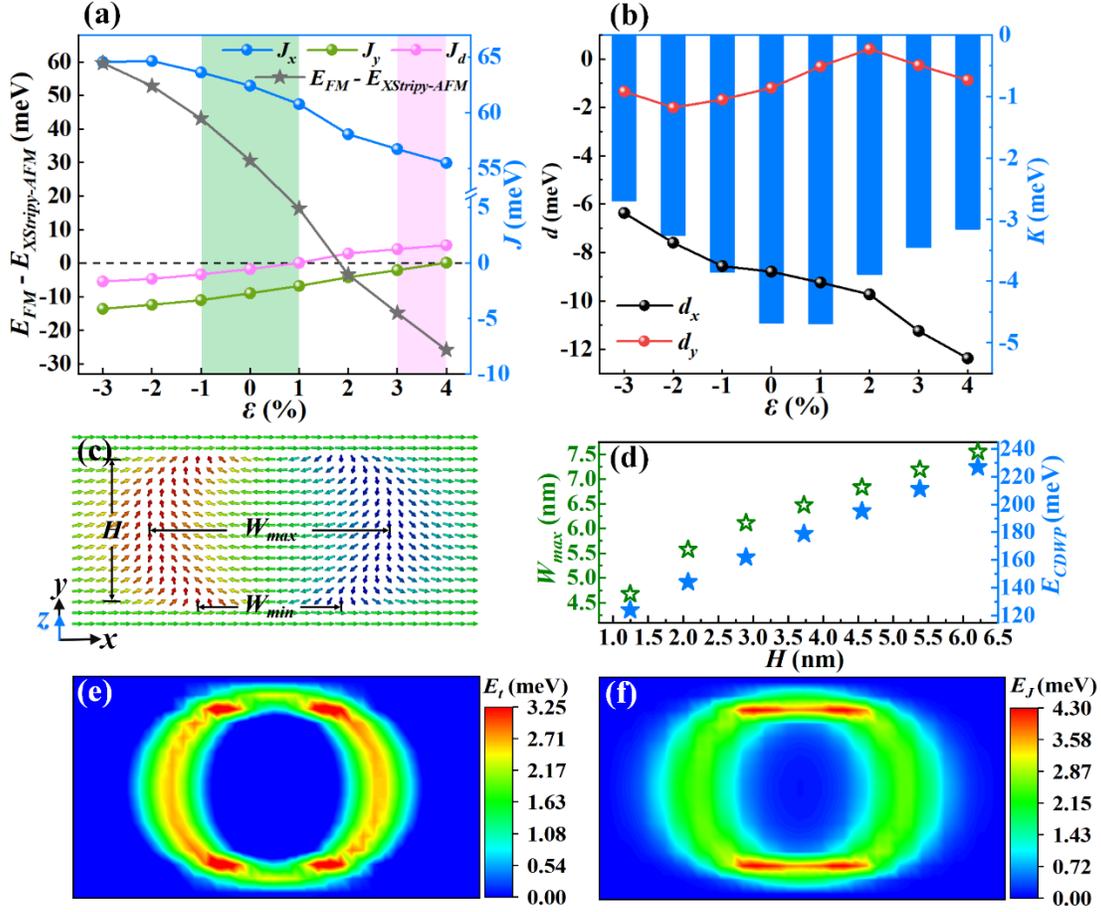

**FIG. 3.** (a) The energy difference between FM and XStripy-AFM states ($E_{FM} - E_{XStripy-AFM}$), the exchange coefficients $J_x$, $J_y$, $J_d$, (b) DMI components $d_x$, $d_y$ and anisotropy coefficient $K$ as a function of biaxial strain $\varepsilon$. (c) The FM CDWP in MnOI monolayer of $\varepsilon = 3\%$. (d) $W_{max}$ and $E_{CDWP}$ of one FM CDWP with different $H$. The distributions of (e) the total energy ($E_t$) and (f) the energy of exchange interactions ($E_J$) for an isolated FM CDWP, where FM state is taken as the reference.

The atomistic simulations show that besides the $\varepsilon$ range from -1% to 1% (light green region) where AFM CDWP exists, there is another $\varepsilon$ range from 3% to 4% (pink region) where the metastable FM CDWP appears on the FM ground state without external magnetic field, as shown in Fig. 3(c). In fact, the AFM CDWP in Fig. 2(b) can be regarded as two nesting FM CDWPs. The profile of FM CDWP, localized in nanoscale, can also be characterized by $W_{max}$, $W_{min}$ and $H$. The FM CDWP is composed of two 180° FM Néel walls with a given chirality, which are tightly bonded with a fixed $W_{min}$ of ~4.3 nm. $W_{max}$ and $E_{CDWP}$ as a function of $H$ show similar behaviors to those of AFM CDWPs, as plotted in Fig. 3(d). The 1D topology with $w = 1$ is maintained by the 360° spin rotation within the $xz$ plane restricted by the strong anisotropy. The crescent shape also originates from the competition between different energy terms, and the exchange interactions play a key role, as shown in Fig. 3(e-f). Thus, the particle-like 2D CDWP with 1D topology in FM version exists in MnOI monolayer under an appropriate strain.

Conventionally, magnetic field is an effective approach to control FM domain walls. Here the dynamics of FM CDWP in MnOI monolayer of $\varepsilon = 3\%$ is investigated under external magnetic field. When the direction of magnetic field is consistent with the spin direction outside the FM CDWP, the CDWP is only suppressed a little but not annihilated by $h$ up to 5.00 T. The robustness of this magnetic texture results from the topological protection. When the magnetic field is applied along the direction of the spins inside the FM CDWP, there is a critical field ($h_c$) that discriminates different dynamic behaviors of FM CDWP. Below $h_c$, its width increases slowly with $h$ increasing, till it approaches a fixed width (Fig. 4(a)). And this fixed width increases exponentially with $h$ rising, as shown in Fig. 4(b). Above $h_c$, the FM CDWP gradually split into a pair of entangled 180° Néel walls, which will be completely separated and move in the opposite directions at a constant speed ($v_h$) (Fig. 4(c)). As plotted in Fig. 4(d), $v_h$ increases linearly with the increasing $h$, and could be at the level of $10^3$ m/s under $h$ of several Teslas.

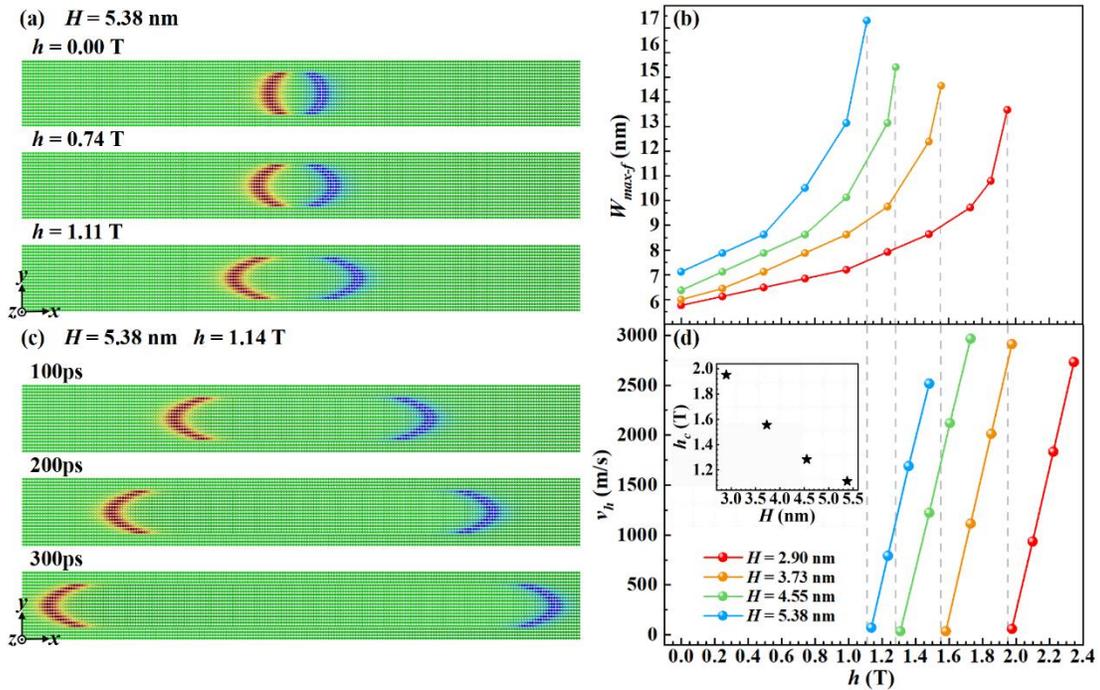

**FIG. 4.** The dynamics of FM CDWP under magnetic field $h$. When $h < h_c$, (a) the snapshots of the FM CDWP ($H = 5.38$ nm) reaching fixed widths under different $h$; (b) the fixed

value of $W_{max}$ ($W_{max\text{-}f}$) as a function of $h$ for FM CDWPs with different $H$. When $h > h_c$, (c) the snapshots of the FM CDWP ($H = 5.38$ nm) under $h = 1.14$ T at 100, 200, 300 ps; (d) the speed ($v_h$) of the separating 180° wall as a function of $h$ for FM CDWPs with different $H$. The inset presents the dependence of $h_c$ on $H$.

These dynamic behaviors under magnetic field also depend on $H$ of FM CDWP. As plotted in Figs. 4(b) and (d), a taller CDWP has a smaller $h_c$, and tends to own a larger $W_{max\text{-}f}$ and a higher speed under the same $h$, and thus is more sensitive to magnetic field. For both cases (below and above $h_c$), if the magnetic field is removed, the elongated CDWP or separated 180° walls will undergo an oscillation process and shrink back to its original state.

## C. Current-driven dynamics

The dynamics of AFM and FM CDMPs driven by in-plane spin-polarized current is simulated by the LLG equation with the term of the spin transfer torque (STT) [36,37,59,60]. The magnitude of STT ($u$) is proportional to the spin-polarized current density (see Note 2(a) in SM [49] for details).

When the spin-polarized current is injected along the $x$ direction, both AFM and FM CDWPs move uniformly in a straight line along the $x$ direction, as displayed in the snapshots of Figs. 5(b-c). The current-driven velocity $v_c$ depends on $u$ linearly, as plotted in Fig. 5(a). FM CDWP is always faster a little than AFM one. Specifically, when $u = 12.00$ m/s (the corresponding current density is estimated to be ~$2.3 \times 10^{11}$ A/m$^2$ approximately) [61], the velocities of both AFM and FM CDWPs could be over 2 km/s, which may be faster than the typical speeds of domain walls or skyrmions under the similar current density. For example, with the driving current at the level of $10^{11}$ A/m$^2$, skyrmion was reported to move at 900 m/s in compensated synthetic antiferromagnets [25]. With the driving current at the level of $10^{12}$ A/m$^2$, it was reported that the domain wall was moved at 750 m/s in synthetic antiferromagnetic racetracks [24], and at 1.3 km/s in compensated ferrimagnets [26].

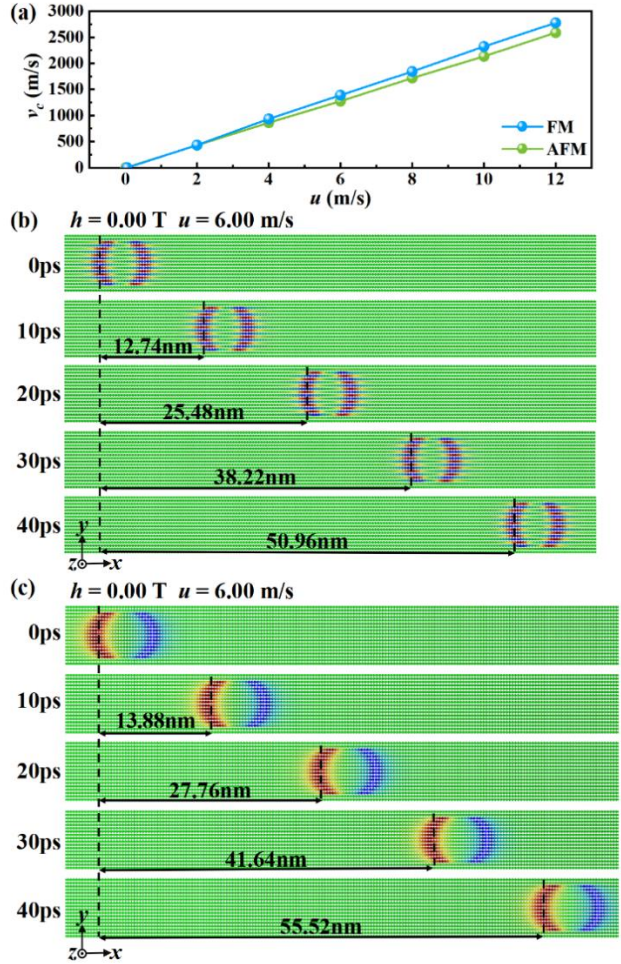

**FIG. 5.** Dynamics of AFM and FM CDWPs driven by current. (a) The current-driven speed $v_c$ as a function of $u$. Snapshots of the moving (b) AFM and (c) FM CDWPs at 0, 10, 20, 30, 40 ps with $u = 6.00$ m/s.

    Due to the strong anisotropy, the motions of both AFM and FM CDWPs are strictly restricted in one dimension, namely along the magnetic easy-axis, no matter what direction of current is applied. When the current is along the $y$ direction, the CDWPs remain motionless. Therefore, CDWP can only be driven by the current component along the magnetic easy-axis. Moreover, the simulation demonstrates that the morphology of CDWP can be well maintained up to higher speed driven by larger current, which confirms its robustness of the particle-like character. Different from the motion driven by magnetic field, the velocity of the current-driven motion does not depend on $H$.

The atomistic simulations at finite temperature show that the CDWPs can survive thermal noise up to 40~50K, and $v_c$ decreases slightly with the increase of temperature (see Note 3(b) in SM [49] for details).

## SUMMARY

In conclusion, by combining first-principles calculations and atomistic simulations, a magnetic monolayer MnOI with strong anisotropy is predicted. Spontaneous AFM CDWP and its FM version are observed in the pristine and strained MnOI monolayer respectively. Originating from the strong anisotropy and the competing interactions, these CDWPs possess topology in one dimension and particle-like robustness in two dimensions, which are different from both conventional 1D domain wall and 2D skyrmion. The unique magnetic structure induces unique dynamic behaviors. The FM CDWP can be disassociated by $h$ higher than $h_c$, and the splitting speed depends on $h$ linearly. Furthermore, both FM and AFM CDWPs can be efficiently driven by spin-polarized current to move rapidly and stably on the 2D plane along a straight line strictly limited by easy-axis anisotropy instead of the specific shape effect (e.g. a nano stripe or track). Thereby, our work provides another possible path to realize the potential application of topological magnetic texture as an information carrier in the design of next-generation high-speed spintronic devices.

## ACKNOWLEDGEMENTS

We thank Ning Ding, Wencong Sun, Ziwen Wang, Xinyu Yang and Lewei Zhou of Southeast University for fruitful discussions. This work is supported by Natural Science Foundation of Jiangsu Province (Grant No. BK20221451), National Natural Science Foundation of China (Grant Nos. 12274070, 124B2064). Most calculations were done on the Big Data Computing Center of Southeast University.